\newcommand*{\ATLASLATEXPATH}{./}
\documentclass[cernpreprint,UKenglish,texlive=2011,txfonts]{\ATLASLATEXPATH atlasdoc}

\usepackage[block=none]{\ATLASLATEXPATH atlaspackage}
\usepackage{multirow}
\usepackage{float}
\usepackage{\ATLASLATEXPATH atlasbiblatex}

\usepackage{\ATLASLATEXPATH atlascontribute}

\usepackage[BSM=true]{\ATLASLATEXPATH atlasphysics}

\graphicspath{{/}{/}}

\usepackage{lldefs}
\usepackage{bm}

\AtlasTitle{Search for high-mass new phenomena in the dilepton final state using proton--proton collisions at $\sqrt{s}~=$~13~\TeV{} with the ATLAS detector}

\usepackage{authblk}
\author{The ATLAS Collaboration%
}

\date{\today}

\AtlasRefCode{EXOT-2015-07}

\PreprintIdNumber{CERN-EP-2016-116}

\AtlasJournal{Phys. Lett. B}
\AtlasJournalRef{Phys. Lett. B 761 (2016) 372-392}
\AtlasDOI{10.1016/j.physletb.2016.08.055}
\AtlasAbstract{%
A search is conducted for both resonant and non-resonant high-mass new phenomena in dielectron and dimuon final states. The search uses \lumill\ of proton--proton collision data, collected at $\sqrt{s}$ = 13~\TeV{} by the ATLAS experiment at the LHC in 2015. The dilepton invariant mass is used as the discriminating variable. No significant deviation from the Standard Model prediction is observed; therefore limits are set on the signal model parameters of interest at 95\% credibility level. Upper limits are set on the cross-section times branching ratio for resonances decaying to dileptons, and the limits are converted into lower limits on the resonance mass, ranging between \ZpllobsLow~\TeV{} and \ZpllobsHigh~\TeV{}, depending on the model. Lower limits on the $\ell\ell qq$ contact interaction scale are set between \CIllobsLow~\TeV{} and \CIllobsHigh~\TeV{}, also depending on the model.}

\AtlasCoverSupportingNote{Supporting documentation}{https://cds.cern.ch/record/2025566}
\AtlasCoverCommentsDeadline{10th of May 2016}

\AtlasCoverAnalysisTeam{G.~Artoni, T.~Berry, D.~Bortoletto, M.~Cano Bret, R.~Brock, J.~Catmore, M.~Chizhov, T.~Dai, E.~Dreyer, F.~Ellinghaus, O.~Fedin, W.~Fedorko, F.~Giuli, D.~Hayden, H.~Herde, H.~Herr, N.~Hod, T.~Hryn’ova, U.~Klein, M.~Levchenko, Y.~Liu, V.~Maleev, P.~Mastrandrea, V.~Morisbak, F.~Ould-Saada, J.~Pearkes, J.~Reichert, S.~Rettie, G.~Savage, S.~Schmitz, G.~Sciolla, I.~Shipsey, B.~Stelzer, O.~Stelzer-Chilton, H.~Torres, C.~Willis, S.~Willocq, Y.~Wu, I.~Yeletskikh, Z.~Zhao, B.~Zhou, M.~Zinser}

\AtlasCoverEdBoardMember{Henri Bachacou~(chair)}
\AtlasCoverEdBoardMember{Sarah Heim}
\AtlasCoverEdBoardMember{Mark Neubauer}

\AtlasCoverEgroupEditors{atlas-exot-2015-07-editors@cern.ch}

\AtlasCoverEgroupEdBoard{atlas-EXOT-2015-07-editorial-board@cern.ch}

\hypersetup{pdftitle={ATLAS draft},pdfauthor={The ATLAS Collaboration}}

\begin{document}

\maketitle



\section{Introduction}
\label{sec:intro}
The dilepton ($ee$ or $\mu\mu$) final-state signature has excellent sensitivity to a wide variety of new phenomena expected in theories beyond the Standard Model (SM). It benefits from high signal selection efficiencies and relatively small, well-understood backgrounds.

Models with extended gauge groups often feature additional \u1\ symmetries
with corresponding heavy spin-1 \zp\ bosons whose decays would manifest themselves as narrow
resonances in the dilepton mass spectrum.
Grand Unified Theories (GUT) have inspired models based on the 
\esix\ gauge group~\cite{London:1986dk, Langacker:2008yv},
which, for a particular choice of symmetry-breaking pattern, includes two neutral gauge 
bosons that mix with an angle \te6. This yields a physical state
defined by $\zp(\te6) = \zppsi\cos\te6 + \zpchi\sin\te6$, where
the gauge fields \zppsi\ and \zpchi\ are associated with two separate \u1\ groups
resulting from the breaking of the \esix\ symmetry. All \zp\ signals in this model are defined by specific values of \te6 ranging from -$\pi$ to $\pi$, and the six commonly motivated cases are investigated in this search, namely \zppsi, \zpeta, \zpN, \zpI, \zpsq, and \zpchi. The widths of these states vary from 0.5\% to 1.2\% of the resonance mass, respectively.
In addition to the GUT-inspired \esix\ models, 
the Sequential Standard Model (SSM)~\cite{Langacker:2008yv}
provides a common benchmark model that includes
a \zpssm\ boson with couplings to fermions identical to those of the SM $Z$ boson.
This search is also sensitive to a series of models that predict the presence of narrow dilepton resonances; however constraints are not explicitly evaluated on these models. These include the Randall--Sundrum (RS) model~\cite{Randall:1999ee} 
with a warped extra dimension giving rise
to spin-2 graviton excitations, 
the quantum black hole model~\cite{Meade:2007sz}, 
the \zstar\ model~\cite{Chizhov:2008tp}, 
and the minimal walking technicolour model~\cite{Sannino:2004qp}.

Some models of physics beyond the SM result in non-resonant deviations 
from the predicted SM dilepton mass spectrum. 
Compositeness models motivated by the repeated pattern of quark and lepton generations
predict new interactions involving their constituents.
These interactions may be represented as a contact interaction (CI)
between initial-state quarks and final-state leptons~\cite{Eichten:1983hw, Eichten:1984eu}.
Other models producing non-resonant effects, but not explicitly evaluated here, are models with large extra dimensions~\cite{ArkaniHamed:1998nn} motivated by the hierarchy problem.
The following four-fermion CI Lagrangian~\cite{Eichten:1983hw, Eichten:1984eu}
is used to describe a new interaction or compositeness in the process 
$q\overline{q} \to \ell^+\ell^-$:

\vspace{-0.4cm}

\begin{eqnarray}\label{lagrangian}
\mathcal L & = & \frac{g^2}{\Lambda^2}\;[ \eta_{\rm LL} \, (\overline q_{\rm L}\gamma_{\mu} q_{\rm L})\,(\overline\ell_{\rm L}\gamma^{\mu}\ell_{\rm L})
 +\eta_{\rm RR} (\overline q_{\rm R}\gamma_{\mu} q_{\rm R}) \,(\overline\ell_{\rm R}\gamma^{\mu}\ell_{\rm R}) \\ \nonumber
 &  & ~~~~~~~~ +\eta_{\rm LR} (\overline q_{\rm L}\gamma_{\mu} q_{\rm L}) \,(\overline\ell_{\rm R}\gamma^{\mu}\ell_{\rm R})
+\eta_{\rm RL} (\overline q_{\rm R}\gamma_{\mu} q_{\rm R}) \,(\overline\ell_{\rm L}\gamma^{\mu}\ell_{\rm L}) ] \: ,
\end{eqnarray}

\noindent where $g$ is a coupling constant set to be $\sqrt{4\pi}$ by convention,
$\Lambda$ is the CI scale,
and $q_{\rm L,R}$ and $\ell_{\rm L,R}$ are left-handed and right-handed quark and lepton fields, respectively. The symbol $\gamma_{\mu}$ denotes the gamma matrices, and the parameters $\eta_{ij}$, where $i$ and $j$ are L or R (left or right), 
define the chiral structure of the new interaction.
Different chiral structures are investigated here, with the left-right (right-left) model
obtained by setting $\eta_{\rm LR} = \pm 1$ ($\eta_{\rm RL} = \pm 1$) and all other parameters to zero.
Likewise, the left-left and right-right models are obtained by setting the corresponding parameters to $\pm 1$, and the others to zero. The sign of $\eta_{ij}$ determines whether the interference
between the SM Drell--Yan (DY) $q\overline{q} \to Z/\gamma^\ast \to \ell^+\ell^-$
process and the CI process is constructive ($\eta_{ij} = -1$) or destructive ($\eta_{ij} = +1$).

The most sensitive previous searches for a \zp\ decaying to the dilepton final state were carried out by the ATLAS and CMS
collaborations~\cite{Aad:2014cka, Khachatryan:2014fba}. Using 20~\ifb{} of $pp$ collision data at $\sqrt{s}$ = 8~\TeV{}, ATLAS set a lower limit at 95\% credibility level (CL) on the \zpssm\ pole mass of 2.90~\TeV{} for the combined $ee$ and $\mu\mu$ channels. Similar limits were set by CMS.
The most stringent constraints on CI searches are also provided by the CMS and ATLAS collaborations~\cite{Aad:2014wca, Khachatryan:2014fba}. The strongest lower limits on the $\ell\ell q q$ CI scale are $\Lambda > 21.6~\TeV{}$ and $\Lambda > 17.2~\TeV{}$ at 95\% CL for constructive and destructive interference, respectively, in the case of left-left interactions and given a uniform positive prior in 1/$\Lambda^2$.
Previous dilepton searches at ATLAS have also set lower limits on the resonance mass in other models such as: an RS graviton up to 2.68~\TeV{}, quantum black holes at 3.65~\TeV{}, the \zstar\ boson at 2.85~\TeV{}, and minimal walking technicolour up to 2.27~\TeV{}~\cite{Aad:2014cka}. Similar lower limits were set by CMS where equivalent searches were performed~\cite{Khachatryan:2014fba}.

In this letter, a search for resonant and non-resonant new phenomena is
presented using the observed $ee$ and $\mu\mu$ mass spectra extracted from
$pp$ collisions within the ATLAS detector at the Large Hadron Collider (LHC) operating at $\rts = 13$ \TeV{}. The $pp$ collision data correspond to an integrated luminosity of \lumill. 
The analysis and interpretation of these spectra rely primarily on simulated samples of 
signal and background processes. 
The $Z$ mass peak region is used to normalise the background contribution
and perform cross-checks of the simulated samples.
The interpretation is then performed taking into account the expected shape of
different signals in the dilepton mass distribution.

\section{ATLAS detector}
\label{sec:detector}
\newcommand{\AtlasCoordFootnote}{%
ATLAS uses a right-handed coordinate system with its origin at the nominal interaction point (IP)
in the centre of the detector and the $z$-axis along the beam pipe.
The $x$-axis points from the IP to the centre of the LHC ring,
and the $y$-axis points upwards.
Cylindrical coordinates $(r,\phi)$ are used in the transverse plane, 
$\phi$ being the azimuthal angle around the $z$-axis.
The pseudorapidity is defined in terms of the polar angle $\theta$ as $\eta = -\ln \tan(\theta/2)$.
Angular distance is measured in units of $\Delta R \equiv \sqrt{(\Delta\eta)^{2} + (\Delta\phi)^{2}}$.}

The ATLAS experiment~\cite{PERF-2007-01, IBL_TDR} at the LHC is a multi-purpose particle detector
with a forward-backward symmetric cylindrical geometry and near $4\pi$ coverage in 
solid angle.\footnote{\AtlasCoordFootnote}
It consists of an inner tracking detector surrounded by a thin superconducting solenoid
providing a \SI{2}{\tesla} axial magnetic field, electromagnetic and hadronic calorimeters, and a muon spectrometer.
The inner tracking detector (ID) covers the pseudorapidity range $|\eta| < 2.5$.
It consists of silicon pixel, silicon microstrip, and transition-radiation tracking detectors.
Lead/liquid-argon (LAr) sampling calorimeters provide electromagnetic (EM) energy measurements
with high granularity.
A hadronic (steel/scintillator-tile) calorimeter covers the central pseudorapidity range ($|\eta| < 1.7$).
The endcap and forward regions are instrumented with LAr calorimeters
for EM and hadronic energy measurements up to $|\eta| = 4.9$. The total thickness of the EM calorimeter is more than twenty radiation lengths.
The muon spectrometer (MS) surrounds the calorimeters and is based on
three large superconducting air-core toroids with eight coils each.
The field integral of the toroids ranges between \num{2.0} and \SI{6.0}{\tesla$\cdot$\metre} for most of the detector.
It includes a system of precision tracking chambers and fast detectors for triggering.
A dedicated trigger system is used to select events.
The first-level trigger is implemented in hardware and uses the calorimeter and muon detectors to reduce the accepted event rate from \SI{40}{\mega\hertz} to below \SI{100}{\kilo\hertz}. This is followed by a software-based trigger that reduces the accepted event rate to \SI{1}{\kilo\hertz} on average.

\section{Data and Monte Carlo samples}
\label{sec:data-mc}
The data sample used in this analysis was collected during the 2015 LHC run with $pp$ collisions at $\sqrt{s}$ = 13~\TeV{}. After selecting periods with stable beams and requiring that relevant detector systems are functional, the data set used for the analysis corresponds to \lumill\ of integrated luminosity. Event quality is also checked to remove those events which contain noise bursts or coherent noise in the calorimeters.

Modelling of the various background sources relies primarily on Monte Carlo (MC) simulation. The dominant background contribution arises from the DY process~\cite{ATL-PHYS-PUB-2016-003}. Other background sources are top-quark~\cite{ATL-PHYS-PUB-2016-004} and diboson ($WW$, $WZ$, $ZZ$)~\cite{ATL-PHYS-PUB-2016-002} production. In the case of the dielectron channel, multi-jet and $W+$jets processes also contribute due to the misidentification of jets as electrons. A data-driven method, described in \refS{sec:background}, is used to estimate these background contributions. The multi-jet and $W+$jets contribution in the dimuon channel is negligible.

DY events are simulated using \powhegbox\ v2~\cite{Alioli:2010xd} at next-to-leading order (NLO) in Quantum Chromodynamics (QCD), and interfaced to the \pythiaeight.186~\cite{Sjostrand:2007gs} parton shower model. The CT10 parton distribution function (PDF) set~\cite{Lai:2010vv} is used in the matrix element calculation. The AZNLO~\cite{AZNLO:2014} set of tuned parameters (``tune'') is used, with the CTEQ6L1 PDF set~\cite{Pumplin:2002vw}, for the modelling of non-perturbative effects. The EvtGen v1.2.0 program~\cite{EvtGen} is used for properties of the bottom and charm hadron decays. \photos {++} version 3.52~\cite{Davidson:2010ew} is used for Quantum Electrodynamic (QED) emissions from electroweak vertices and charged leptons. Event yields are corrected with a mass-dependent rescaling to next-to-next-to-leading order (NNLO) in the QCD coupling constant, computed with \vrap~0.9~\cite{Anastasiou:2003ds} and the CT14NNLO PDF set~\cite{Dulat:2015mca}. The NNLO QCD corrections are a factor of $\sim$ 0.98 at $m_{\ell\ell}$ = 3~\TeV{}. Mass-dependent electroweak (EW) corrections are computed at NLO with \mcsanc~1.20~\cite{Bondarenko:2013nu}. The NLO EW corrections are a factor of $\sim$ 0.86 at $m_{\ell\ell}$ = 3~\TeV{}. Those include photon-induced contributions ($\gamma\gamma \to \ell\ell$ via $t$- and $u$-channel processes) computed with the MRST2004QED PDF set~\cite{Martin:2004dh}.

Diboson processes with four charged leptons, three charged leptons and one neutrino, or two charged leptons and two neutrinos are simulated using the \sherpa\ 2.1.1 generator~\cite{Gleisberg:2008ta}. Matrix elements contain all diagrams with four electroweak vertices. They are calculated for up to one (4$\ell$, 2$\ell$+2$\nu$) or no additional partons (3$\ell$+1$\nu$) at NLO. Diboson processes with one of the bosons decaying hadronically and the other leptonically are simulated using the \sherpa\ 2.1.1 generator. They are calculated for up to one ($ZZ$) or no ($WW$, $WZ$) additional partons at NLO. All are calculated with up to three additional partons at leading-order (LO) using the Comix~\cite{Gleisberg:2008fv} and OpenLoops~\cite{Cascioli:2011va} matrix element generators and merged with the \sherpa\ parton shower~\cite{Schumann:2007mg} using the ME+PS@NLO prescription~\cite{Hoeche:2012yf}. The CT10 PDF set is used in conjunction with dedicated parton shower tuning developed by the \sherpa\ authors. The \sherpa\ diboson sample cross-section was scaled down to account for its use of \alphaqed=1/129 rather than 1/132 corresponding to the use of current PDG parameters as input to the $G_{\mu}$ scheme.

For the generation of \ttbar\ and single top quarks in the $Wt$-channel and $s$-channel the \powhegbox\ v2 generator with the CT10 PDF set in the matrix element calculations is used. EW $t$-channel single-top-quark events are generated using the \powhegbox\ v1 generator. This generator uses the four-flavour scheme for the NLO matrix element calculations together with the fixed four-flavour PDF set CT10f4. For all top-quark processes, top-quark spin correlations are preserved (for $t$-channel, top quarks are decayed using MadSpin~\cite{Artoisenet:2012st}). The parton shower, fragmentation, and the underlying event are simulated using \pythiasix.428~\cite{Sjostrand:2006za} with the CTEQ6L1 PDF set and the Perugia 2012 tune (P2012)~\cite{Skands:2010ak}. The top-quark mass is set to 172.5~\GeV{}. The EvtGen v1.2.0 program is used for properties of the bottom and charm hadron decays. The \ttbar\ and single-top-quark MC samples are normalised to a cross-section as calculated with the Top++ 2.0 program~\cite{Czakon:2011xx}, which is accurate to NNLO in perturbative QCD, including resummation of next-to-next-to-leading logarithmic soft gluon terms.

Resonant and non-resonant signal processes are produced at LO using \pythiaeight.186 with the NNPDF23LO PDF set~\cite{Ball:2012cx} and A14 tune~\cite{ATL-PHYS-PUB-2014-021} for event generation, parton showering and hadronisation. In the case of \zp\ production, interference effects (such as with DY production) are not included. However, for the production of non-resonant signal events, both the DY and CI events are generated together in the same sample to account for the significant interference effects between those two processes. Higher-order QCD corrections are computed as for the DY background and applied to both the resonant and non-resonant MC samples. EW corrections are not applied to the resonant MC samples due to the large model dependence. However, these corrections are applied to the non-resonant MC samples as they involve interference between the DY and CI processes. Moreover, including the EW corrections leads to a more conservative estimate when setting exclusion limits. The generator settings and corrections described here are also used to compute the signal cross-sections and branching ratios.

The detector response is simulated with \geant~4~\cite{Agostinelli:2002hh, Aad:2010ah} and the events are processed with the same reconstruction software as used for the data. Furthermore, the distribution of the number of additional simulated $pp$ collisions in the same or neighbouring beam crossings (pile-up) is accounted for by overlaying simulated minimum-bias events and re-weighting the MC to match the distribution observed in the data.

\section{Event selection}
\label{sec:selection}

Electrons are reconstructed in the central region of the ATLAS detector covered by the tracking detectors ($|\eta|<\mathrm{2.47}$), by combining calorimetric and tracking information as described in Ref.~\cite{ATLAS-CONF-2014-032}. The transition region between the central and forward regions of the calorimeters, in the range 1.37 $\le$ |$\eta$| $\le$ 1.52, exhibits degraded energy resolution and is therefore excluded. A likelihood discriminant is built to suppress electron candidates resulting from hadronic jets, photon conversions, Dalitz decays and semileptonic heavy-flavour hadron decays. The likelihood discriminant utilises lateral and longitudinal shower shape, tracking and cluster--track matching quantities. Several operating points are defined for the likelihood discrimination, as described in Ref.~\cite{ATLAS-CONF-2014-032}. In this analysis, the \textit{Medium} working point is used in the search, and the \textit{Very Loose} and \textit{Loose} working points are used in the data-driven background estimation described in \refS{sec:background}. In addition to the likelihood discriminant, selection criteria based on track quality are applied. The selection efficiency smoothly decreases from 96\% to 95\% for electrons with transverse energy (\et) between 500~\GeV{} and 1.5~\TeV{}. The selection efficiency modelling is evaluated in the data using a tag-and-probe method~\cite{ATLAS-CONF-TNP} up to \et\ of 500~\GeV{} and the uncertainties due to the modelling of the shower shape variables are evaluated as described in Section~\ref{sec:systematics}. The electron energy scale and resolution has been calibrated up to \et\ of 500~\GeV{} using data taken at $\sqrt{s}$ = 8~\TeV{}~\cite{Aad:2014nim}. The energy resolution for high-\et\ electrons is approximately 1\%.
To suppress background from misidentified jets as well as from light- and heavy-flavour hadron decays inside jets, electrons are required to satisfy the calorimeter-based and track-based isolation criteria with a fixed efficiency of 99\% over the full range of electron momentum. The calorimeter-based isolation relies on the ratio of the total energy deposited in a cone of size $\Delta R = 0.2$ centred at the electron cluster barycentre to the electron \et. Likewise, the track-based isolation relies on the ratio of the scalar sum of transverse momenta of tracks within a cone of size $\Delta R = 10\GeV/\pt$ to the transverse momentum (\pt) of the electron track. The tracks are required to originate from the primary vertex (defined as the vertex with the highest sum of track $\pt^{2}$), have $\pt > 1 \GeV$, $|\eta| < 2.5$, and meet track quality criteria. 
 

Candidate muon tracks are, at first, reconstructed independently in the ID and the MS~\cite{MuonPerf}. 
The two tracks are then used as input to a combined fit which takes into account the energy loss in the calorimeter and multiple-scattering effects.
The ID track used for the combined fit is required to be within the ID acceptance, $|\eta|<2.5$, and to have a minimum number of hits in each ID sub-system.
Muon candidates in the overlap of the MS barrel and endcap region ($1.01<|\eta|<1.10$) are rejected due to the potential for \pt\ mismeasurement resulting from relative barrel--endcap misalignment.
In order to reduce the background from light- and heavy-hadron decays inside jets, muons are required to fulfil relative track-based isolation requirements with a fixed efficiency of 99\%,
as defined above for electron candidates. 
The selected muon candidates must also pass near the primary interaction point in the $z$ coordinate to suppress cosmic-ray background. 
Since momentum resolution is a key ingredient of this analysis, 
muon tracks are required to have at least three hits in each of three precision chambers in the MS and not to traverse regions of the MS which are poorly aligned.
This requirement reduces the muon reconstruction efficiency by about 20\% for muons with a \pt\ greater than $1.5\TeV$.
Finally, the \qp\ (charge divided by momentum) measurements performed independently in the ID and MS must agree within seven standard deviations, calculated
from the sum in quadrature of the ID and MS momentum uncertainties.

To search for high-mass dilepton signatures of new physics, requirements are applied to the data and MC samples to select events with two high-\et\ electrons or high-\pt\ muons, satisfying the criteria described above. In the dielectron channel, a two-electron trigger based on the \textit{Loose} identification criteria with an \et\ threshold of 17~\GeV{} for each electron is used. Events in the dimuon channel are required to pass at least one of two single-muon triggers with \pt{} thresholds of 26~\GeV{} and 50~\GeV{}, with the former also requiring the muon to be isolated. These triggers select events from a simulated sample of \zpchi\ with a pole mass of 3~\TeV{} with an efficiency of about 87\% and 94\% for the dielectron and dimuon channels, respectively. Electron (muon) candidates are required to have \et (\pt) greater than 30~\GeV{} and have a transverse impact parameter consistent with the beam-line. Events are required to have at least one reconstructed primary vertex and at least one pair of same-flavour lepton candidates. 

Only the electron (muon) pair with the highest scalar sum of \et\ (\pt) is retained in each event
and an opposite-charge requirement is applied in the dimuon case. The opposite-charge requirement is not applied in the dielectron channel due to higher chance of charge misidentification for high-\et\ electrons.

Energy (momentum) calibration and resolution smearing are applied to electron (muon) candidates in the simulated samples to match the performance observed in data~\cite{Aad:2014nim, MuonPerf}. Event-level corrections are applied in the simulated samples to match the trigger, reconstruction and isolation efficiencies.

Representative values of the total acceptance times efficiency for a \zpchi\ boson with a pole mass of 3~\TeV{} are 69\% in the dielectron channel and 46\% in the dimuon channel.

\section{Background estimation}
\label{sec:background}

The backgrounds from processes producing two real leptons in the final state are modelled using MC simulated samples as described in \refS{sec:data-mc}. The processes for which MC simulation is used are: DY, $t\bar{t}$ and single-top-quark, and diboson ($WW$, $WZ$, and $ZZ$) production. The simulated samples for the top-quark (single and pair) production and diboson production are not large enough to model the dilepton mass distribution above several hundred \GeV. Therefore, fits to the dilepton invariant mass spectrum (\mll{}) using monotonically decreasing functions are used to extrapolate these background processes to dilepton masses above 600~\GeV{}.

In the dimuon channel, contributions from $W+$jets and multi-jet production are negligible, and therefore are not included in the expected yield. However, the $W+$jets, multi-jet and other production processes, where at most one real electron is produced, do contribute to the selected $ee$ sample due to their having one or more hadronic jets satisfying the electron selection criteria. The contribution from these processes is estimated simultaneously with a data-driven technique, the \textit{matrix method}, described in Ref.~\cite{Aad:2014cka}. In this technique, probabilities for electrons and jets to pass electron candidate selection are used.  Probabilities of electron identification are estimated from MC simulated DY samples in several bins of \et\ and $|\myeta|$. Probabilities of jet misidentification as an electron in different \et\ bins are estimated in data samples triggered on the presence of a \textit{Very Loose} or a \textit{Loose} electron candidate.
The estimate is extrapolated by fitting a smooth function to the \mee{} distribution between 150 and 600~\GeV{} to mitigate effects of limited event counts in the high-mass region and method instability in the $Z$ peak region. The uncertainties in this background estimate are evaluated by considering differences in the estimates for events with same-charge and opposite-charge electrons as well as by varying the electron identification probabilities and changing the parameters of the extrapolation functions.

As a final step, the sums of backgrounds estimated using MC samples are rescaled independently in both channels so that the estimated count of events matches the data in the $Z$-peak normalisation region $\mathrm{80} \GeV < \mll < \mathrm{120} \GeV$. This normalisation procedure is found to agree with the equivalent scaling using the expected integrated luminosity within 2\% for both channels (and in the same direction), which is well within the current luminosity uncertainty of 5\%. The luminosity uncertainty was calculated using the same methodology as for the 7~\TeV{} data~\cite{Aad:2013ucp}.

\section{Systematic uncertainties}
\label{sec:systematics}
As a result of the background yield normalisation described above, the background prediction is insensitive to the luminosity uncertainty as well as any other mass-independent effect. Signal scaling is performed using the event counts in the data in the Z-peak region. Therefore, a uniform uncertainty of 4\% due to the uncertainty in the $Z/\gamma^*$ cross-section in the normalisation region is applied to signal. This uncertainty was obtained using a calculation based on \vrap\ at NNLO evaluating the effect of varying the PDF sets, scales and \alphas{}. Mass-dependent systematic uncertainties, on the other hand, are considered as nuisance parameters in the statistical interpretation and include both the theoretical and experimental effects on the total background and experimental effects on the signal. Systematic uncertainties common to the dielectron and dimuon channels are treated as correlated where relevant. All systematic uncertainties estimated to have an impact $<$ 3\% on the total expected number of events for all values of $m_{\ell\ell}$ are neglected, as they have a negligible impact on the results of the search.

Theoretical uncertainties in the background prediction are dominated by the DY background in this search. They arise from the PDF eigenvector variations of the nominal PDF set, as well as variations of PDF scale, \alphas, EW corrections, and photon-induced (PI) corrections. The effects of different PDF set choices are also considered. The theoretical uncertainties are the same at generator level for the dielectron and dimuon channels, but result in different uncertainties at reconstruction level, due to the differing resolutions between the two channels. The PDF variation uncertainty is obtained using the 90\% CL CT14NNLO PDF error set and by following the procedure described in Refs.~\cite{Aad:2014cka, Gao:2013bia, Butterworth:2015oua}. Rather than using a single nuisance parameter to describe the 28 eigenvectors of this PDF error set, which could lead to an underestimation of its effect, a re-diagonalised set of 7 PDF eigenvectors was used~\cite{Dulat:2015mca}, which are treated as separate nuisance parameters. The sum in quadrature of these eigenvectors matches the original CT14NNLO error envelope well. The uncertainties due to the variation of PDF scale and \alphas\ are derived using \vrap\ with the former obtained by varying the renormalisation and factorisation scales of the nominal CT14NNLO PDF up and down simultaneously by a factor of two. The value of \alphas\ used (0.118) is varied by $\pm$ 0.003. The EW correction uncertainty was assessed by comparing the nominal additive (1+$\delta_{\mathrm{EW}}$+$\delta_{\mathrm{QCD}}$) treatment with the multiplicative approximation ((1+$\delta_{\mathrm{EW}}$)(1+$\delta_{\mathrm{QCD}}$)) treatment of the EW correction in the combination of the higher-order EW and QCD effects. The uncertainty in the photon-induced correction is calculated based on the uncertainty of the quark masses and the photon PDF. An additional uncertainty is derived due to the choice of nominal PDF set, by comparing the central values of CT14NNLO with those from other PDF sets as recommended by the PDF4LHC forum~\cite{Butterworth:2015oua}, namely MMHT14~\cite{Motylinski:2014sya} and NNPDF3.0~\cite{Ball:2014uwa}. 
The maximum absolute deviation from the envelope of these comparisons is used as the PDF choice uncertainty, where it is larger than the CT14NNLO PDF eigenvector variation envelope. Theoretical uncertainties are not applied to the signal prediction in the statistical interpretation.

Theoretical uncertainties in the $\ttbar$ and diboson backgrounds were also considered. The $\ttbar$ MC sample is normalised to a cross-section of $\sigma_{\ttbar}$ = $832^{+20}_{-29}$ (scale) $\pm$ 35 (PDF + $\alpha_{\mathrm{S}}$) pb, calculated with the Top++ 2.0 program as described in \refS{sec:data-mc}. The first uncertainty comes from the independent variation of the factorisation and renormalisation scales, $\mu_F$ and $\mu_R$, while the second one is associated to variations in the PDF and $\alpha_{\mathrm{S}}$, following the PDF4LHC prescription~\cite{Butterworth:2015oua}. Normalisation uncertainties in the top quarks and diboson background were found to be negligible. The uncertainties in the top-quark and diboson background extrapolations are estimated by varying both the functional form and the fit range, taking the envelope of all variations. These uncertainties were also found to be negligible with respect to the total background estimate. Both sources of systematic uncertainty in these background contributions are included in the ``Top quarks \& dibosons'' entry in Table~\ref{systematics}.

\begin{table}
  \begin{center}
    \caption{Summary of the relative systematic uncertainties in the expected number of events at a dilepton mass of 2~\TeV~(3~\TeV). The background estimate is normalised to data in the dilepton invariant mass window 80--120~\GeV{}, and the values quoted for the uncertainty represent the relative change in the total expected number of events in the given \mll\ histogram bin containing the reconstructed \mll\ mass of 2~\TeV~(3~\TeV). For the signal uncertainties the values were computed using a \zpchi\ signal model with a pole mass of 2~\TeV~(3~\TeV) by comparing yields in the core of the mass peak (within the full width at half maximum) between the distribution varied by a given uncertainty and the nominal distribution. The total uncertainty quoted on the last line is obtained from a sum in quadrature of the individual uncertainties. ``\na" represents cases where the uncertainty is not applicable, and ``negligible'' represents cases where the uncertainty is smaller than 3\% across the entire mass spectrum, which are neglected in the statistical interpretation. \label{systematics}}
    \centering
     \begin{tabular}{l|cc|cc}
      \toprule
      \toprule
      Source  &  \multicolumn{2}{c|}{Dielectron}  &  \multicolumn{2}{c}{Dimuon}  \\
      &  Signal  &  Background  &  Signal  &  Background  \\
      \midrule
      Normalisation               & \syspair{4.0}{4.0}                       & \na                                       & \syspair{4.0}{4.0}                       & \na  \\
      PDF choice                  & \na                                      & \syspairb{\lessthan{1.0}}{\lessthan{1.0}} & \na                                      & \syspair{\lessthan{1.0}}{\lessthan{1.0}}  \\
      PDF variation               & \na                                      & \syspairb{9.1}{13.5}                      & \na                                      & \syspair{8.2}{11.1}  \\
      PDF scale                   & \na                                      & \syspairb{1.8}{2.3}                       & \na                                      & \syspair{1.7}{2.0}  \\  
      \alphas\                    & \na                                      & negligible                                & \na                                      & negligible \\
      EW corrections              & \na                                      & \syspairb{2.3}{3.9}                       & \na                                      & \syspair{2.0}{3.1} \\
      Photon-induced corrections  & \na                                      & \syspairb{3.4}{5.4}                       & \na                                      & \syspair{3.1}{4.3}   \\
      Top quarks \& dibosons      & \na                                      & negligible                                & \na                                      & negligible \\  
      Efficiency                  & \syspair{5.4}{5.4}                       & \syspairb{5.4}{5.4}                       & \syspair{13.6}{17.6}                     & \syspair{13.6}{17.6} \\
      Lepton scale \& resolution  & \syspair{\lessthan{1.0}}{\lessthan{1.0}} & \syspairb{3.7}{5.4}                       & \syspair{4.7}{4.8}                       & \syspair{2.3}{6.9} \\
      Multi-jet \& $W$+jets       & \na                                      & negligible                                & \na                                      & \na \\
      MC statistical              & negligible                               & negligible                                & negligible                               & negligible \\  
      \midrule
      Total                       & \syspair{6.7}{6.7}                       & \syspairb{12.1}{17.0}                     & \syspair{14.9}{18.7}                     & \syspair{16.6}{22.6} \\
      \bottomrule
      \bottomrule
    \end{tabular}
  \end{center}
\end{table}

The following sources of experimental uncertainty are accounted for: lepton trigger, identification, reconstruction, and isolation efficiency, lepton energy scale and resolution, multi-jet and $W$+jets background estimate, and MC statistics. Efficiencies are evaluated using events from the $Z \to \ell\ell$ peak and then extrapolated to high energies. The uncertainty in the muon reconstruction efficiency is the largest experimental uncertainty in the dimuon channel. It includes the uncertainty obtained from $Z \to \mu\mu$ data studies and a high-\pt\ extrapolation uncertainty corresponding to the magnitude of the decrease in the muon reconstruction and selection efficiency with increasing \pt\ that is predicted by the MC simulation. The effect on the muon reconstruction efficiency was found to be approximately 3\% per \TeV{} as a function of muon \pt. The uncertainty in the electron identification efficiency extrapolation is based on the differences in the electron shower shapes in the EM calorimeters between data and MC simulation in the $Z \to ee$ peak, which are propagated to the high-\et\ electron sample. The effect on the electron identification efficiency was found to be 2.0\% and is independent of \et\ for electrons with \et\ above 150~\GeV{}. Mismodelling of the muon momentum resolution due to residual misalignments in the MS can alter the steeply falling background shape at high dilepton mass and can significantly modify the width of the signal line shape. This uncertainty is obtained by studying dedicated data-taking periods with no magnetic field in the MS~\cite{MuonPerf}. For the dielectron channel, the uncertainty includes a contribution from the multi-jet and $W$+jets data-driven estimate that is obtained by varying both the overall normalisation and the extrapolation methodology, which is explained in \refS{sec:background}. Systematic uncertainties used in the statistical analysis of the results are summarised in Table~\ref{systematics} at dilepton mass values of 2~\TeV{} and 3~\TeV{}.

\section{Event yields}
\label{sec:yields}
Expected and observed event yields, in bins of invariant mass, are shown in~\refT{yields_table} for the dielectron (top) and dimuon (bottom) channels. Expected event yields are split into the different background sources and the yields for two signal scenarios. The DY process is dominant over the entire mass range.
In general, the observed data are in good agreement with the SM prediction, taking uncertainties into account as described in \refS{sec:systematics}. 
A deficit is observed for the dimuon channel in the invariant mass region between 300~\GeV{} and 500~\GeV{}. Extensive cross-checks were performed in this region, and the deficit was quantified by calculating the local Poisson $p$-value, using the sources of systematic uncertainty described in \refS{sec:systematics}, which gives a significance less than two standard deviations for this mass interval. 
 
\begin{table}[!h]
  \caption{Expected and observed event yields in the dielectron (top) and dimuon (bottom) channels in different dilepton mass intervals. 
  The quoted errors for the dominant Drell--Yan background correspond to the combined statistical, theoretical, and experimental systematic uncertainties. The errors quoted for the other background sources correspond to the combined statistical and experimental systematic uncertainties.}
  \label{yields_table}
  \centering
  \tiny
  \resizebox{\textwidth}{!}{%
  \begin{tabular}{l r@{ $\pm$ }l r@{ $\pm$ }l r@{ $\pm$ }l r@{ $\pm$ }l r@{ $\pm$ }l r@{ $\pm$ }l r@{ $\pm$ }l r@{ $\pm$ }l }
    \toprule
    \toprule
    \mee\ [\GeV] & \multicolumn{2}{c}{120--300} & \multicolumn{2}{c}{300--500} & \multicolumn{2}{c}{500--700} & \multicolumn{2}{c}{700--900} & \multicolumn{2}{c}{900--1200} & \multicolumn{2}{c}{1200--1800} & \multicolumn{2}{c}{1800--3000} & \multicolumn{2}{c}{3000--6000} \\
    \midrule 
    Drell--Yan ($Z$/$\gamma^*$) &  21000 & 400 & 940 & 50 & 149 & 10 & 38.3 & 3.0 & 16.5 & 1.4 & 5.6 & 0.6 & 0.78 & 0.10 & 0.030 & 0.005 \\
    Top Quarks                  &  4550 & 110 & 446 & 25 & 47.2 & 1.6 & 6.2 & 0.8 & 1.13 & 0.35 & 0.12 & 0.09 & 0.002 & 0.006 & \lesspermil \\
    Diboson                     &  620 & 10 & 67.5 & 1.2 & 10.3 & 0.9 & 2.3 & 0.5 & 0.78 & 0.28 & 0.20 & 0.11 & 0.021 & 0.018 & \lesspermil \\
    Multi-Jet \& $W$+Jets       &  320 & 80 & 40 & 12 &  7.2   & 1.8  & 1.6  & 0.8  & 0.5  & 0.4  & 0.08 & 0.10 & 0.002 & 0.005 & \lesspermil \\
    \midrule 
    Total SM                    &  26500 & 400 & 1490 & 60 &  214   & 11   & 48.4   & 3.2    & 18.9 & 1.6  & 6.0  & 0.6  & 0.81  & 0.10  & 0.030 & 0.006 \\
    \midrule 
    Data                     & \multicolumn{2}{c}{25951} & \multicolumn{2}{c}{1447} & \multicolumn{2}{c}{202} & \multicolumn{2}{c}{44} & \multicolumn{2}{c}{17} & \multicolumn{2}{c}{9} & \multicolumn{2}{c}{0} & \multicolumn{2}{c}{0} \\ 
    \midrule 
    SM+\zp\ ($m_{\zp}$ = 3 \TeV{})           &  26500 & 400 & 1490 & 60 &  214 & 11 & 48.4 & 3.2 & 19.0 & 1.6 & 6.0 & 0.6 & 2.3 & 0.5 & 0.9 & 0.5 \\
    SM+CI ($\Lambda^{\rm Const}_{\rm LL}$ = 20 \TeV{}) &  26500 & 400 & 1500 & 60 &  220 & 11 & 52.1 & 3.2 & 22.2 & 1.6 & 8.8 & 0.6 & 2.22 & 0.14 & 0.289 & 0.018 \\
    \midrule
    \midrule
    \mmm\ [\GeV] & \multicolumn{2}{c}{120--300} & \multicolumn{2}{c}{300--500} & \multicolumn{2}{c}{500--700} & \multicolumn{2}{c}{700--900} & \multicolumn{2}{c}{900--1200} & \multicolumn{2}{c}{1200--1800} & \multicolumn{2}{c}{1800--3000} & \multicolumn{2}{c}{3000--6000} \\
    \midrule 
    Drell--Yan ($Z$/$\gamma^*$)           & 19300 & 400 & 770 & 31 & 115 & 7 & 29.0 & 2.2 & 11.8 & 1.0 & 4.0 & 0.4 & 0.61 & 0.09 & 0.034 & 0.007 \\
    Top Quarks                            & 3855 & 29 & 369 & 9 & 43.4 & 2.5 & 7.5 & 0.5 & 1.97 & 0.16 & 0.36 & 0.04 & 0.020 & 0.004 & \lesspermil \\
    Diboson                               & 412.1 & 3.4 & 43.7 & 0.9 & 7.08 & 0.30 & 1.67 & 0.11 & 0.61 & 0.05 & 0.174 & 0.023 & 0.020 & 0.006 & \lesspermil \\
    \midrule 
    Total SM                              & 23600 & 400 & 1183 & 32 & 165 & 7 & 38.1 & 2.2 & 14.4 & 1.0 & 4.6 & 0.4 & 0.65 & 0.09 & 0.036 & 0.008 \\
    \midrule 
    Data       & \multicolumn{2}{c}{23275} & \multicolumn{2}{c}{1083} & \multicolumn{2}{c}{164} & \multicolumn{2}{c}{29} & \multicolumn{2}{c}{13} & \multicolumn{2}{c}{5} & \multicolumn{2}{c}{0} & \multicolumn{2}{c}{0} \\ 
    \midrule 
    SM+\zp\ ($m_{\zp}$ = 3 \TeV{})           & 23600 & 400 & 1183 & 32 & 165 & 7 & 38.1 & 2.2 & 14.4 & 1.0 & 4.6 & 0.4 & 1.27 & 0.12 & 0.55 & 0.09 \\
    SM+CI ($\Lambda^{\rm Const}_{\rm LL}$ = 20 \TeV{})  &  23600 & 400 & 1193 & 32 & 174 & 8 & 41.9 & 2.4 & 16.8 & 1.2 & 6.4 & 0.6 & 1.49 & 0.20 & 0.164 & 0.028 \\
    \bottomrule
    \bottomrule
  \end{tabular}}
\end{table}

Distributions of \mll{} in the dielectron and dimuon channels are shown in Fig.~\ref{plots_inv_mass}. No significant excess is observed. The highest invariant mass event is found at 1775~\GeV{} in the dielectron channel, and 1587~\GeV{} in the dimuon channel. Both of these events appear to be very clean with little other detector activity, apart from an accompanying jet in the dimuon event.

\begin{figure}[!htb]
  \subfloat[][]{
    \includegraphics[width=0.49\linewidth]{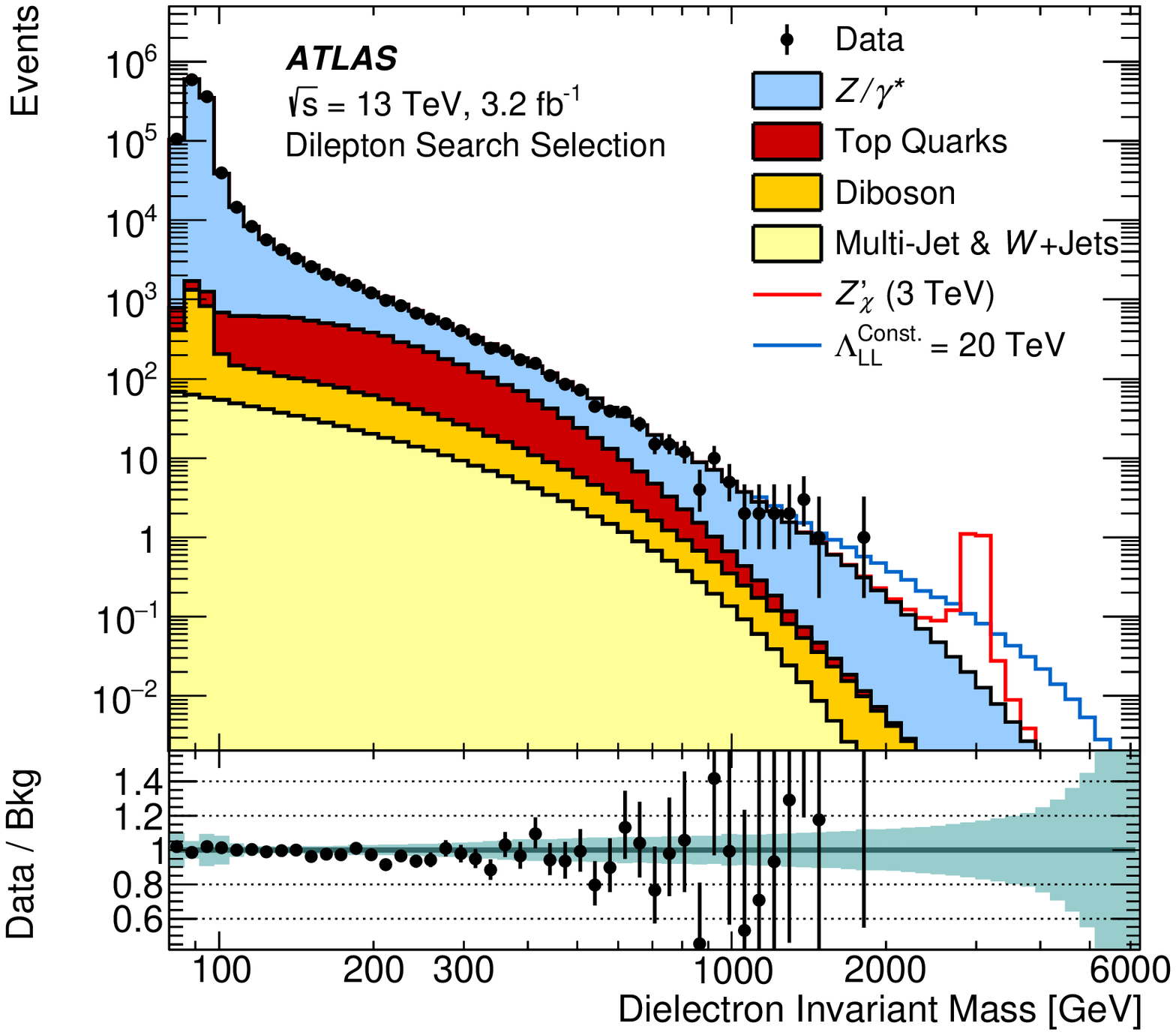}\label{plots_inv_mass_ee}
  }
  \subfloat[][]{
    \includegraphics[width=0.49\linewidth]{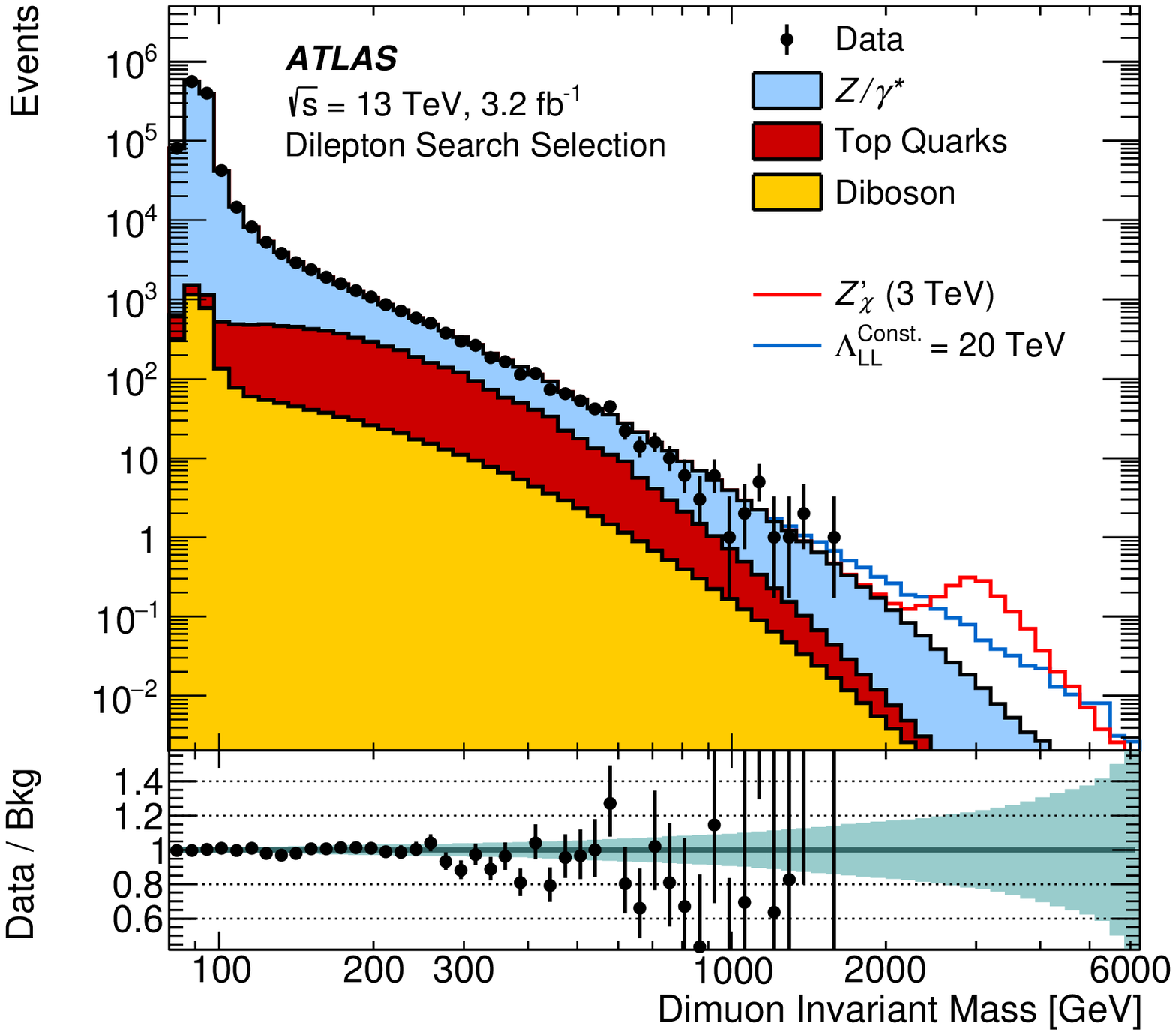}\label{plots_inv_mass_mm}
  }
  \caption{Distributions of \protect\subref{plots_inv_mass_ee} dielectron and \protect\subref{plots_inv_mass_mm} dimuon 
  reconstructed invariant mass (\mll) after selection, for data and the SM background estimates as well as their ratio. 
  Two selected signals are overlaid; resonant \zpchi\ with a pole mass of 3~\TeV\ and 
  non-resonant contact interactions with left-left (LL) constructive interference and $\Lambda$ = 20~\TeV. 
  The bin width of the distributions is constant in log(\mll), and the shaded band 
  in the lower panel illustrates the total systematic uncertainty, as explained in~\refS{sec:systematics}. The data points are shown together with their statistical uncertainty.}
  \label{plots_inv_mass}
\end{figure}

\section{Statistical analysis}
\label{sec:statistics}

A search for a resonant signal is performed using the \mll{} distribution in the dielectron and dimuon channels utilising the log-likelihood ratio (LLR) test described in \citeR{stat_asymptotic}. To perform the LLR search, the HistFactory~\cite{Cranmer:1456844} package, together with RooStats~\cite{RooStats} and RooFit~\cite{Verkerke:2003ir} packages are used. The $p$-value for finding a \zpchi\ signal excess (at a given pole mass) more significant than the observed, is computed analytically using a test statistic based on the logarithm of the profile likelihood ratio $\lambda(\mu)$ which includes a treatment of the systematic uncertainties. The parameter $\mu$ is defined as a ratio of the signal production cross-section times branching ratio to the dilepton final state (\xbr) to its theoretically predicted value. The test statistic is modified for signal masses below 800~\GeV{} to also quantify the significance of potential deficits in the data. The analytical calculation of $p$-values is cross-checked using MC simulations. Multiple mass hypotheses are tested in pole-mass steps corresponding to the histogram bin width to compute the local $p$-values --- that is $p$-values corresponding to specific signal mass hypotheses. The chosen bin width for the \mll{} histogram corresponds to the resolution in the dielectron (dimuon) channel, which varies from 10 (60)~\GeV{} at \mll\ = 1~\TeV{} to 15 (200)~\GeV{} at \mll\ = 2~\TeV{}, and 20 (420)~\GeV{} at \mll\ = 3~\TeV{}. Pseudo-experiments are used to estimate the distribution of the lowest local $p$-value in the absence of any signal. The $p$-value to find anywhere in the \mll{} distribution (120--6000~\GeV{}) an excess more significant than the one in the data (global $p$-value) is then computed.
The {\sc BumpHunter} method~\cite{Choudalakis:2011qn} is also used in a model-independent search for an excess in all consecutive intervals in the \mll\ histogram spanning from one bin to half of the bins in the histogram. The same binning as in the LLR search is used.

Upper limits on the \zp\ \xbr\ and lower limits on the CI scale $\Lambda$ in a variety of interference and chiral coupling scenarios are set in a Bayesian approach. The logarithmic \mll\ histogram binning shown in Fig.~\ref{plots_inv_mass} uses 66 mass bins and is chosen for setting limits on resonant signals using \zpchi\ signal templates. For setting the limits on the CI interaction scale $\Lambda$, the \mll{} mass distribution uses eight mass bins above 400~\GeV{} with bin widths varying from 100 to 1500~\GeV{}.
The prior probability is chosen to be uniform and positive in the cross-section for the $\zp$ limit calculation and $1/\Lambda^2$ or $1/\Lambda^4$ for the CI limit calculation. For the CI limit calculation, these choices of the prior are selected to study the cases where the dilepton production cross-section is dominated by the interference terms and where it is dominated by the pure contact interaction term. The upper (lower) 95\% percentile of the posterior probability is then quoted as the upper (lower) 95\% credibility-level limit on \xbr\ ($\Lambda$). The above calculations are performed with the Bayesian Analysis Toolkit (BAT)~\cite{Caldwell:2008fw}, which uses a Markov Chain MC technique to integrate over the nuisance parameters. Limit values obtained using the experimental data are quoted as observed limits, while median values of the limits from a large number of pseudo-experiments, where only SM background is present, are quoted as the expected limits. The upper limits on the \xbr\ in a \zp\ model are interpreted as lower limits on the \zp\ pole mass using the relationship between the pole mass and the theoretical \zp\ cross-section.

\section{Results}
\label{sec:result}
The statistical tests described in the previous section do not reveal a signal. The LLR tests for a \zpchi\ find global $p$-values of \LLRpvalee{}, \LLRpvalmm\, and \LLRpvalcomb\ in the dielectron, dimuon, and combined channels, respectively. The {\sc BumpHunter}~\cite{Choudalakis:2011qn} test, which scans the mass spectrum with varying intervals to find the most significant excess in data, finds $p$-values of \BHpvalee\ and \BHpvalmm\ in the dielectron and dimuon channels, respectively. The largest deviation from the background-only hypothesis using the LLR tests for a \zpchi\ is observed at 192~\GeV\ in the dimuon mass spectrum with a local significance of $2.5\,\sigma$, but is not globally significant ($0.6\,\sigma$). There are also smaller but noticeable excesses in dimuon channel at 583~\GeV\ with a local significance of $1.8\,\sigma$, in the dielectron channel at 652~\GeV\ with a local significance of $2.0\,\sigma$, and in the combined dilepton channel at 1410~\GeV\ with a local significance of $2.0\,\sigma$.
Upper limits on the cross-section times branching ratio (\xbr) for \zp\ bosons are presented in Fig.~\ref{fig-limits-Zp}. The observed and expected lower pole-mass limits for various \zp\ scenarios are summarised in \refT{tab-limits-zp}. The upper limits on \xbr\ for \zp\ bosons start to weaken above a pole mass of $\sim$~3~\TeV. This is mainly due to the combined effect of a rapidly-falling signal cross-section as the kinematic limit is approached, and the natural width of the resonance. The effect is more pronounced in the dimuon channel due to worse mass resolution than in the dielectron channel. The selection efficiency also starts to slowly decrease at very high pole mass, but this is a sub-dominant effect. The lower limits on the CI scale, $\Lambda$, where a prior uniform and positive in 1/$\Lambda^2$ is used are summarised in Fig.~\ref{fig-limits-CI}. \refT{tab-limits-ci} gives an overview of $\Lambda$ lower limits for all considered chiral coupling and interference scenarios as well as both choices of the prior $\Lambda$ probability.

\begin{figure}[h]
\begin{center}
\subfloat[]{\includegraphics[width=0.52\linewidth]{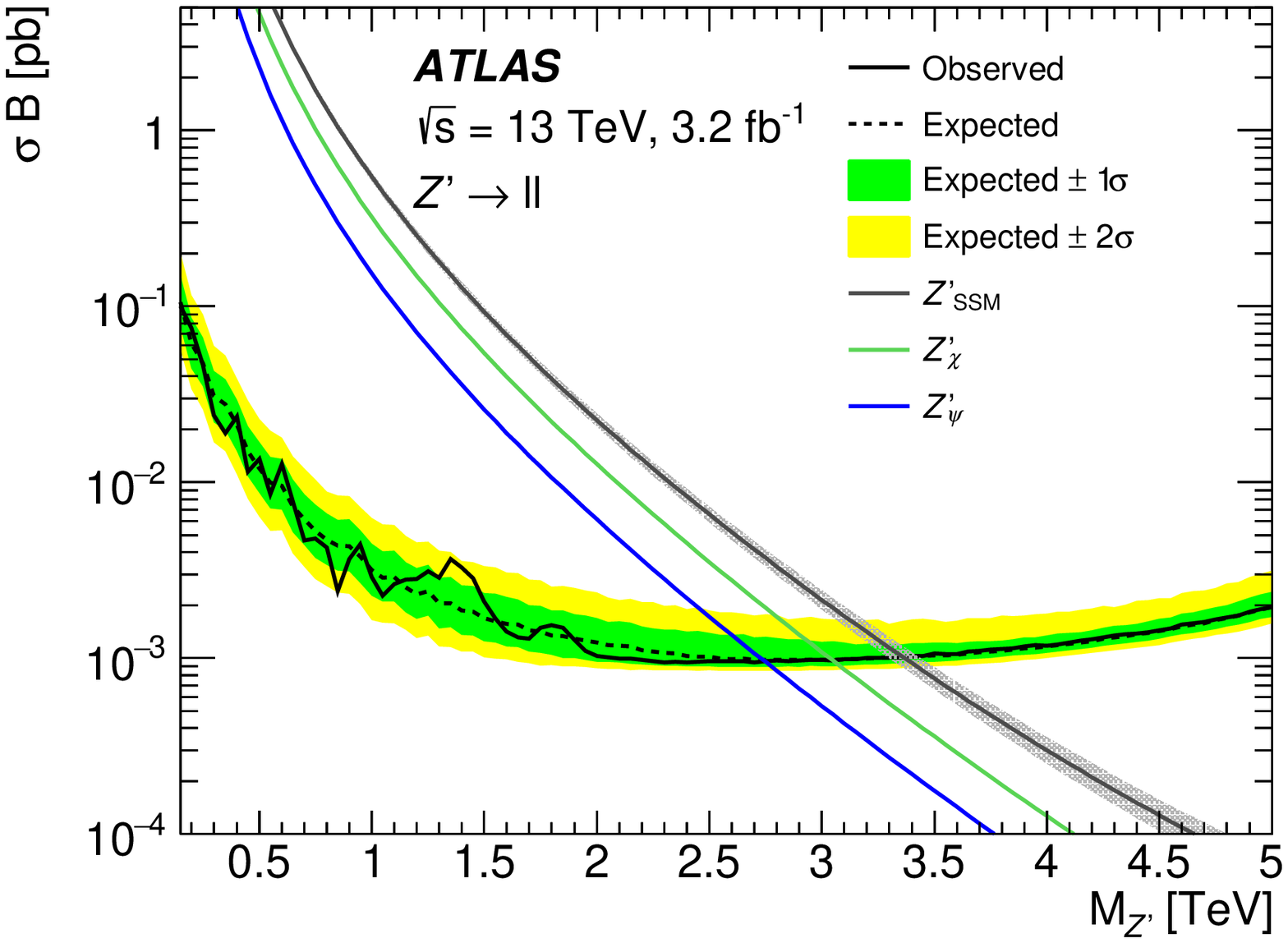}\label{fig-limits-Zp}}
\subfloat[]{\includegraphics[width=0.52\linewidth]{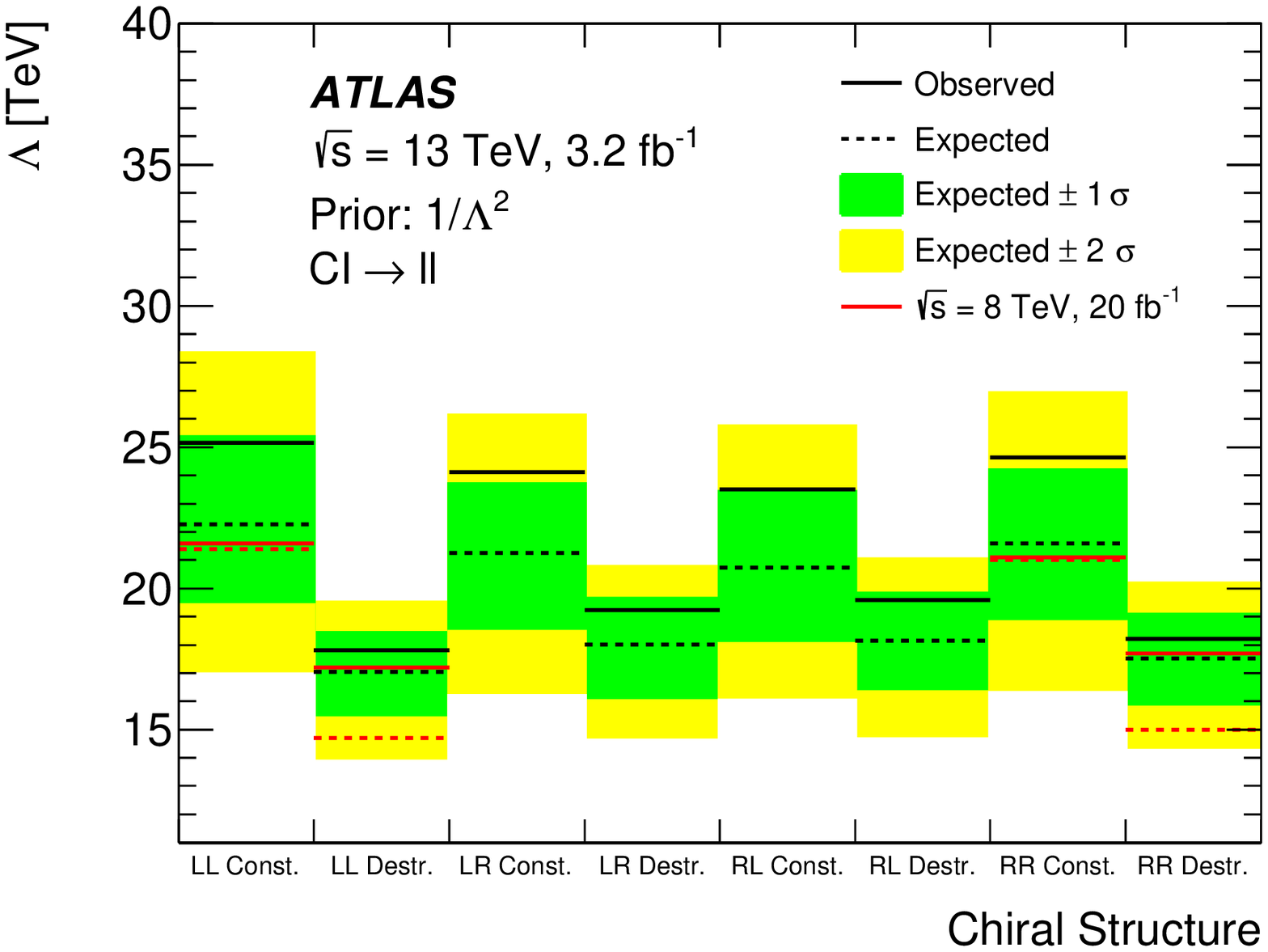}\label{fig-limits-CI}}
\end{center}
\caption{(a) Upper 95\% CL limits for \zp\ production cross-section times branching ratio to two leptons as a function of \zp\ pole mass (M$_{\zp}$). The signal theory lines are calculated with \pythia~8 using the NNPDF23LO PDF set~\cite{Ball:2012cx}, and corrected to next-to-next-to-leading order in QCD using \vrap~\cite{Anastasiou:2003ds} and the CT14NNLO PDF set~\cite{Dulat:2015mca}. The signal theoretical uncertainties are shown as a band on the \zpssm\ theory line for illustration purposes, but are not included in the \xbr\ limit calculation. (b) Lower 95\% CL limits on the contact interaction (CI) scale $\Lambda$ for different chiral couplings and both constructive and destructive interference scenarios using a uniform positive prior in 1/$\Lambda^2$. For the left-left (LL) and right-right (RR) cases, the ATLAS $\rts = 8$ \TeV{} results~\cite{Aad:2014wca} are shown for comparison. In that publication, the left-right (LR) case was obtained by setting $\eta_{\rm LR} = \eta_{\rm RL} = \pm 1$ and therefore is not directly comparable to the results presented here.}
\end{figure}

\begin{table}[h]
  \begin{center}
    \caption{Observed and expected 95\% CL lower mass limits for various \zp\ gauge boson models. The widths are quoted as a percentage of the resonance mass.
    \label{tab-limits-zp}}
    \centering
    \addtolength{\tabcolsep}{+1pt}
    \vspace{0.2cm}
    \begin{tabular}{|c|c|r@{.}l|cc|cc|cc|}
      \hline
      \hline
      \multirow{3}{*}{Model} & \multirow{3}{*}{Width [\%]} & \multicolumn{2}{c|}{}    & \multicolumn{6}{c|}{Lower limits on $m_{\zp}$ [\TeV{}]} \\
\cline{5-10} 
                             &                             & \multicolumn{2}{c|}{\te6 [rad]}  & \multicolumn{2}{c|}{$ee$} & \multicolumn{2}{c|}{$\mu\mu$} & \multicolumn{2}{c|}{$\ell\ell$} \\
\cline{5-10}
                             &                             & \multicolumn{2}{c|}{}   & Obs & Exp &                    Obs & Exp &                   Obs & Exp \\
      \hline
      \zpssm\ & 3.0 & \multicolumn{2}{c|}{-}           & \zpeessmobs\ & \zpeessmexp\ & \zpmmssmobs\ & \zpmmssmexp\ & \zpllssmobs\ & \zpllssmexp\ \\
      \zpchi\ & 1.2 & ~0 & 50~$\pi$           & \zpeechiobs\ & \zpeechiexp\ & \zpmmchiobs\ & \zpmmchiexp\ & \zpllchiobs\ & \zpllchiexp\ \\
      \zpsq\  & 1.2 & ~0 & 63~$\pi$ & 2.83 & 2.81 & 2.54 & 2.57 & 3.00 & 3.00 \\   
      \zpI\   & 1.1 & ~0 & 71~$\pi$  & 2.77 & 2.76 & 2.49 & 2.51 & 2.94 & 2.94 \\ 
      \zpeta\ & 0.6 & ~0 & 21~$\pi$  & 2.63 & 2.62 & 2.35 & 2.36 & 2.81 & 2.80 \\ 
      \zpN\   & 0.6 & ~-0 & 08~$\pi$  & 2.63 & 2.62 & 2.35 & 2.36 & 2.80 & 2.80 \\           
      \zppsi\ & 0.5 & \multicolumn{2}{c|}{0} & \zpeepsiobs\ & \zpeepsiexp\ & \zpmmpsiobs\ & \zpmmpsiexp\ & \zpllpsiobs\ & \zpllpsiexp\ \\
      \hline
      \hline
    \end{tabular}
  \end{center}
\end{table}

\begin{table}[h]
\newcommand\T{\rule{0pt}{2.6ex}}
\newcommand\B{\rule[-1.2ex]{0pt}{0pt}}
\caption{Observed and expected 95\% CL lower limits on $\Lambda$ for the LL, LR, RL, and RR chiral coupling scenarios, for both the constructive and destructive interference cases using a uniform positive prior in 1/$\Lambda^2$ or 1/$\Lambda^4$. The dielectron, dimuon, and combined dilepton channel limits are shown.}
\label{tab-limits-ci}
\vspace{0.2cm}
\centering
\begin{tabular}{|c|c|c|c|c|c|c|c|c|c|}
\hline
\hline
\multirow{3}{*}{Channel} & \multirow{3}{*}{Prior} & \multicolumn{8}{c|}{Lower limits on $\Lambda$ [\TeV{}]} \T\B \\
\cline{3-10}
  & & \multicolumn{2}{c|}{Left-Left} \T & \multicolumn{2}{c|}{Left-Right} & \multicolumn{2}{c|}{Right-Left} & \multicolumn{2}{c|}{Right-Right} \\
\cline{3-10}
& & Const. \T & Destr. & Const. & Destr. & Const. & Destr. & Const. & Destr. \\
\hline
\hline
Obs: $ee$ \T & \multirow{2}{*}{1/$\Lambda^2$} & 19.5 & 15.5 & 18.7 & 16.2 & 18.5 & 16.4 & 18.5 & 16.4 \\
Exp: $ee$ &  & 19.5 & 15.8 & 18.7 & 16.5 & 18.4 & 16.5 & 18.4 & 16.6 \\
\hline
Obs: $ee$ \T & \multirow{2}{*}{1/$\Lambda^4$} & 17.7 & 14.4 & 17.0 & 15.0 & 16.8 & 15.1 & 16.8 & 15.1 \\
Exp: $ee$ &  & 17.6 & 14.7 & 16.9 & 15.3 & 16.8 & 15.3 & 16.8 & 15.4 \\
\hline \hline
Obs: $\mu\mu$ \T & \multirow{2}{*}{1/$\Lambda^2$} & 21.8 & 15.8 & 21.1 & 16.9 & 20.5 & 17.2 & 22.0 & 15.7 \\
Exp: $\mu\mu$ &  & 17.9 & 14.5 & 17.4 & 15.2 & 17.2 & 15.3 & 17.9 & 14.5 \\
\hline
Obs: $\mu\mu$ \T & \multirow{2}{*}{1/$\Lambda^4$} & 19.0 & 14.9 & 18.5 & 15.7 & 18.1 & 15.9 & 19.1 & 14.8 \\
Exp: $\mu\mu$ &  & 16.5 & 13.9 & 16.1 & 14.5 & 15.9 & 14.5 & 16.7 & 13.9 \\
\hline \hline
Obs: $\ell\ell$ \T & \multirow{2}{*}{1/$\Lambda^2$}  & 25.2 & 17.8 & 24.1 & 19.2 & 23.5 & 19.6 & 24.6 & 18.2 \\
Exp: $\ell\ell$ &  & 22.3 & 17.0 & 21.3 & 18.0 & 20.7 & 18.1 & 21.6 & 17.5 \\
\hline
Obs: $\ell\ell$ \T & \multirow{2}{*}{1/$\Lambda^4$} & 22.2 & 16.7 & 21.3 & 17.8 & 21.0 & 18.1 & 21.7 & 17.0 \\
Exp: $\ell\ell$ &  & 20.2 & 15.9 & 19.6 & 17.0 & 19.1 & 17.0 & 19.5 & 16.5 \\
\hline
\hline
\end{tabular}
\end{table}


\FloatBarrier

\section{Conclusions}
\label{sec:conclusion}
The ATLAS detector at the Large Hadron Collider has been used to search for resonant and non-resonant new phenomena in the dilepton invariant mass spectrum above the $Z$-boson pole. The search is conducted with \lumill\ of $pp$ collision data at $\sqrt{s}$ = 13~\TeV{}, recorded during 2015. The highest invariant mass event is found at 1775~\GeV{} in the dielectron channel, and 1587~\GeV{} in the dimuon channel. The observed dilepton invariant mass spectrum is consistent with the Standard Model prediction, within systematic and statistical uncertainties. Among a choice of different models, the data are interpreted in terms of resonant spin-1 \zp\ gauge boson production and non-resonant contact interactions. Upper limits are therefore set on the cross-section times branching ratio for a spin-1 \zp\ gauge boson. The resulting lower mass limits are \zpllssmobs~TeV for the \zpssm, \zpllchiobs~\TeV{} for the \zpchi, and \zpllpsiobs~\TeV{} for the \zppsi. Other \esix\ \zp\ models are also constrained in the range between those quoted for the \zpchi\ and \zppsi. These are more stringent than the previous ATLAS result obtained at $\sqrt{s}$ = 8~\TeV{}, by up to 450~\GeV{}. The lower limits on the energy scale $\Lambda$ for various $\ell\ell qq$ contact interaction models range between \CIllobsLow~\TeV{} and \CIllobsHigh~\TeV{}, which are more stringent than the previous ATLAS result obtained at $\sqrt{s}$ = 8~\TeV{}, by up to 3.6~\TeV{}.


\clearpage

\section*{Acknowledgements}

We thank CERN for the very successful operation of the LHC, as well as the
support staff from our institutions without whom ATLAS could not be
operated efficiently.

We acknowledge the support of ANPCyT, Argentina; YerPhI, Armenia; ARC, Australia; BMWFW and FWF, Austria; ANAS, Azerbaijan; SSTC, Belarus; CNPq and FAPESP, Brazil; NSERC, NRC and CFI, Canada; CERN; CONICYT, Chile; CAS, MOST and NSFC, China; COLCIENCIAS, Colombia; MSMT CR, MPO CR and VSC CR, Czech Republic; DNRF and DNSRC, Denmark; IN2P3-CNRS, CEA-DSM/IRFU, France; GNSF, Georgia; BMBF, HGF, and MPG, Germany; GSRT, Greece; RGC, Hong Kong SAR, China; ISF, I-CORE and Benoziyo Center, Israel; INFN, Italy; MEXT and JSPS, Japan; CNRST, Morocco; FOM and NWO, Netherlands; RCN, Norway; MNiSW and NCN, Poland; FCT, Portugal; MNE/IFA, Romania; MES of Russia and NRC KI, Russian Federation; JINR; MESTD, Serbia; MSSR, Slovakia; ARRS and MIZ\v{S}, Slovenia; DST/NRF, South Africa; MINECO, Spain; SRC and Wallenberg Foundation, Sweden; SERI, SNSF and Cantons of Bern and Geneva, Switzerland; MOST, Taiwan; TAEK, Turkey; STFC, United Kingdom; DOE and NSF, United States of America. In addition, individual groups and members have received support from BCKDF, the Canada Council, CANARIE, CRC, Compute Canada, FQRNT, and the Ontario Innovation Trust, Canada; EPLANET, ERC, FP7, Horizon 2020 and Marie Sklodowska-Curie Actions, European Union; Investissements d'Avenir Labex and Idex, ANR, R{\'e}gion Auvergne and Fondation Partager le Savoir, France; DFG and AvH Foundation, Germany; Herakleitos, Thales and Aristeia programmes co-financed by EU-ESF and the Greek NSRF; BSF, GIF and Minerva, Israel; BRF, Norway; Generalitat de Catalunya, Generalitat Valenciana, Spain; the Royal Society and Leverhulme Trust, United Kingdom.

The crucial computing support from all WLCG partners is acknowledged gratefully, in particular from CERN, the ATLAS Tier-1 facilities at TRIUMF (Canada), NDGF (Denmark, Norway, Sweden), CC-IN2P3 (France), KIT/GridKA (Germany), INFN-CNAF (Italy), NL-T1 (Netherlands), PIC (Spain), ASGC (Taiwan), RAL (UK) and BNL (USA), the Tier-2 facilities worldwide and large non-WLCG resource providers. Major contributors of computing resources are listed in Ref.~\cite{ATL-GEN-PUB-2016-002}.
\clearpage
\pagebreak
\sloppy
\printbibliography

\newpage
\begin{flushleft}
{\Large The ATLAS Collaboration}

\bigskip

M.~Aaboud$^\textrm{\scriptsize 136d}$,
G.~Aad$^\textrm{\scriptsize 87}$,
B.~Abbott$^\textrm{\scriptsize 114}$,
J.~Abdallah$^\textrm{\scriptsize 65}$,
O.~Abdinov$^\textrm{\scriptsize 12}$,
B.~Abeloos$^\textrm{\scriptsize 118}$,
R.~Aben$^\textrm{\scriptsize 108}$,
O.S.~AbouZeid$^\textrm{\scriptsize 138}$,
N.L.~Abraham$^\textrm{\scriptsize 150}$,
H.~Abramowicz$^\textrm{\scriptsize 154}$
% ...
\bigskip
\\
% ...
$^{12}$ Institute of Physics, Azerbaijan Academy of Sciences, Baku, Azerbaijan\\
% ...
$^{65}$ University of Iowa, Iowa City IA, United States of America\\
% ...
$^{87}$ CPPM, Aix-Marseille Universit{\'e} and CNRS/IN2P3, Marseille, France\\
% ...
$^{108}$ Nikhef National Institute for Subatomic Physics and University of Amsterdam, Amsterdam, Netherlands\\
% ...
$^{114}$ Homer L. Dodge Department of Physics and Astronomy, University of Oklahoma, Norman OK, United States of America\\
% ...
$^{118}$ LAL, Univ. Paris-Sud, CNRS/IN2P3, Universit{\'e} Paris-Saclay, Orsay, France\\
% ...
$^{136}$ $^{(a)}$ Facult{\'e} des Sciences Ain Chock, R{\'e}seau Universitaire de Physique des Hautes Energies - Universit{\'e} Hassan II, Casablanca; $^{(b)}$ Centre National de l'Energie des Sciences Techniques Nucleaires, Rabat; $^{(c)}$ Facult{\'e} des Sciences Semlalia, Universit{\'e} Cadi Ayyad, LPHEA-Marrakech; $^{(d)}$ Facult{\'e} des Sciences, Universit{\'e} Mohamed Premier and LPTPM, Oujda; $^{(e)}$ Facult{\'e} des sciences, Universit{\'e} Mohammed V, Rabat, Morocco\\
% ...
$^{138}$ Santa Cruz Institute for Particle Physics, University of California Santa Cruz, Santa Cruz CA, United States of America\\
% ...
$^{150}$ Department of Physics and Astronomy, University of Sussex, Brighton, United Kingdom\\
% ...
$^{154}$ Raymond and Beverly Sackler School of Physics and Astronomy, Tel Aviv University, Tel Aviv, Israel\\
% ...
$^{a}$ Also at Department of Physics, King's College London, London, United Kingdom\\
$^{b}$ Also at Institute of Physics, Azerbaijan Academy of Sciences, Baku, Azerbaijan\\
$^{c}$ Also at Novosibirsk State University, Novosibirsk, Russia\\
$^{d}$ Also at TRIUMF, Vancouver BC, Canada\\
$^{e}$ Also at Department of Physics {\&} Astronomy, University of Louisville, Louisville, KY, United States of America\\
% ...
\end{flushleft}


\end{document}